# Uncovering an Interfacial Band Resulting from Orbital Hybridization in Nickelate Heterostructures

Mingyao Chen,[1,#] Huimin Liu,[1,#] Xu He,[2,#] Minjuan Li,[1,#] Chi Sin Tang,[1,3,*] Mengxia Sun,[1] Krishna Prasad Koirala,[4] Mark E. Bowden,[4] Yangyang Li,[5] Xiongfang Liu,[1] Difan Zhou,[1] Shuo Sun,[1] Mark B.H. Breese,[3,6] Chuanbing Cai,[1] Yingge Du,[4] Andrew T. S. Wee,[6,7] Le Wang,[4,*] Xinmao Yin[1,*]

[1]Shanghai Key Laboratory of High Temperature Superconductors, Department of Physics, Shanghai University, Shanghai 200444, China

[2]Theoretical Materials Physics, Q-MAT, CESAM, Université de Liège, B-4000 Liège, Belgium

[3]Singapore Synchrotron Light Source (SSLS), National University of Singapore, Singapore 117603

[4]Physical and Computational Sciences Directorate, Pacific Northwest National Laboratory, Richland, WA 99354, USA

[5]School of Physics, Shandong University, Jinan, 250100, Shandong, China

[6]Department of Physics, Faculty of Science, National University of Singapore, Singapore 117542

[7]Centre for Advanced 2D Materials and Graphene Research, National University of Singapore, Singapore 117546

*Corresponding author

slscst@nus.edu.sg (C.S.T.)

le.wang@pnnl.gov (L.W)

yinxinmao@shu.edu.cn (X.Y.)

**Abstract**

The interaction of atomic orbitals at the interface of perovskite oxide heterostructures has been investigated for its profound impact on the band structures and electronic properties, giving rise to unique electronic states and a variety of tunable functionalities. In this study, we conducted an extensive investigation of the optical and electronic properties of epitaxial NdNiO$_3$ thin films grown on a series of single crystal substrates. Unlike films synthesized on other substrates, NdNiO$_3$ on SrTiO$_3$ (NNO/STO) gives rise to a unique band structure which features an additional unoccupied band situated above the Fermi level. Our comprehensive investigation, which incorporated a wide array of experimental techniques and density functional theory calculations, revealed that the emergence of the interfacial band structure is primarily driven by the orbital hybridization between Ti *3d* orbitals of the STO substrate and O *2p* orbitals of the NNO thin film. Furthermore, exciton peaks have been detected in the optical spectra of the NNO/STO film, attributable to the pronounced electron-electron (e-e) and electron-hole (e-h) interactions propagating from the STO substrate into the NNO film. These findings underscore the substantial influence of interfacial orbital hybridization on the electronic structure of oxide thin-films, thereby offering key insights into tuning their interfacial properties.

## INTRODUCTION

Strongly-correlated oxide systems have emerged as a central focus in the realm of condensed matter physics, largely due to the complex many-body interactions inherent within these systems [1, 2]. Interfaces in oxides, with unique structural characteristics, offer a platform for exploring novel phenomena not observable in bulk materials [3, 4]. Moreover, significant advances in device technology have underscored the significance of interfaces in these materials [5, 6], emphasizing their pivotal role in leveraging the unique properties of strongly correlated oxides within this field.

Interfacial electronic states can undergo significant changes influenced by factors, such as orbital hybridization, lattice mismatches from strain, and charge transfer. In first-row transition metal oxides, the hybridization between the $3d$ band of transition metal component and the $2p$ band of oxygen component significantly influences their electronic and band properties. Previous studies have emphasized the crucial role of interfacial hybridization in governing charge transfer dynamics between nickelates and other perovskite oxides [7, 8, 9], as well as its impact on charge localization in copper-based high-temperature superconductors [10]. In addition, interfacial hybridization has been observed to significantly affect the magnetic properties [11, 12, 13, 14, 15], leading to a variety of other quantum effects [16, 17, 18]. Therefore, elucidating the mechanism and effects of interfacial orbital hybridization is essential for a more thorough and comprehensive understanding of material properties.

However, the investigation of specific impacts of the interface within oxide heterostructures remains challenging due to their complex nature and diverse structure variations. In addressing this challenge, perovskite oxide systems, characterized by their unique structural and electronic features, offer an ideal platform for such inquiries [19, 20, 21]. The rotational and distortive dynamics of oxygen octahedra in perovskite oxides play a crucial role in inducing asymmetry at the oxide/substrate interface. This asymmetry, in turn, influences the spin-orbit coupling and gives rise to various interfacial phenomena [22, 23, 24]. Among the perovskite oxides, rare earth nickelates ($RNiO_3$, where R denotes Lanthanide) have attracted increasing attention for their intriguing electronic properties and the emergence of superconductivity [25, 26, 27]. The structural and physical properties of nickelates can be modified by adjusting the rare earth element [28, 29, 30, 31]. Furthermore, extensive studies on the interfacial effects of nickelates have unveiled intriguing phenomena absent in bulk materials [32, 33, 34, 35, 36].

In this study, we employed spectroscopic techniques to probe the interfacial electronic structure of $NdNiO_3$ films grown on different substrates. Our comprehensive experimental approach, including temperature-dependent spectroscopic ellipsometry and X-ray absorption spectroscopy (XAS), yielded a detailed understanding of the optical spectra associated with these heterostructures. Our results reveal clear changes in electronic structures when varying the substrate, including the appearance of a new high-energy state and the detection of exciton peaks. Detailed density functional theory (DFT) calculations further support our experimental findings, revealing a strong hybridization effect between Ti $3d$ orbitals of the STO substrates and O $2p$ orbitals of the NNO film at the NNO/STO interface. This intricate interfacial hybridization process critically contributes to the radical changes observed in the electronic structures of the NNO film. These findings highlight the significance of interfacial orbital hybridization, offering valuable methodologies and insights for investigating the interfacial effects in complex oxide heterostructures.

## RESULTS and DISCUSSIONS

### Synthesis and characterization of NNO films

High-quality epitaxial NNO thin films with thicknesses between 4.0 and 12.0 nm were deposited on (001)-oriented $LaAlO_3$ (LAO), $(LaAlO_3)_{0.3}(Sr_2AlTaO_6)_{0.7}$ (LSAT), and STO substrates using pulsed laser deposition (PLD). NNO possesses an orthorhombic crystal structure with a pseudocubic room temperature lattice constant of ~3.81 Å in its bulk form [37, 38], which is larger than that of the LAO substrate ($a_{LAO} \approx 3.793$ Å), but smaller than that of LSAT ($a_{LSAT} \approx 3.868$ Å) and STO ($a_{STO} \approx 3.905$ Å) substrates. Hence, the NNO film experiences a compressive strain of ~-0.47% when deposited on LAO and a tensile strain of 1.52% (2.49%) when grown on the LSAT (STO) substrate, as shown in Fig. 1a. Fig. 1b displays the X-ray diffraction (XRD) $\theta$-$2\theta$ scans near the (001) peaks for the representative 12.0 nm NNO films. The strain effect induces a notable change in the out-of-plane lattice parameter of NNO films, as evidenced by the peak shift to higher 2θ angle when transitioning from LAO to STO. The distinct thickness fringes around the Bragg peaks indicate the high-quality of NNO films. Reciprocal space maps (RSM) (Fig. 1c) near the (103) reflection confirm that the 12.0 nm thick NNO films are coherently strained to the substrates. Fig. 1d presents a high-angle annular dark-field scanning transmission electron microscopy (HAADF-STEM) image of the 12 nm thick NNO film on STO. This image shows the good quality of the sample and reveals the absence of apparent structural defects, such as Ruddlesden-Popper phases and NiO secondary phases, within the NNO film. Furthermore, atomic-resolution energy-dispersive X-ray spectroscopy (EDS) maps were conducted to unveil the chemical distribution across the NNO/STO interface. The EDS maps for Nd (yellow), Ni (blue), Sr (red), and Ti (green) elements are presented individually. A superimposed overlay of these maps confirms a stacking sequence of $NiO_2/NdO/TiO_2/SrO$ at the interface, with some Ti/Ni intermixing due to Ti out-diffusion. This observation aligns with recent findings regarding the polar interface of infinite-layer nickelate thin films [39].

**Electrical transport properties and optical conductivity spectra of NNO films**

Fig. 2a shows the temperature-dependent resistivity of NNO films grown on STO with varying thicknesses. With increasing the NNO film thickness from 4.0 to 12.0 nm, $T_{MI}$ gradually rises and approaches the bulk value of 200 K. However, for 12.0 nm thick NNO films, there is an obvious increase in the resistivity and the metal-insulator transition appears to be less distinct. This behavior could be attributed to the presence of oxygen vacancies, possibly induced by the larger tensile strain effect [40, 41, 42]. The temperature-dependent resistivity of NNO on other substrates is shown in Fig. S1. Further analysis of the carrier dynamics can also be seen in our optical measurements derived from the spectroscopic ellipsometry. This will be discussed in further detail later.

The distinctive electrical transport properties of NNO films on different substrates and with different thicknesses are readily reflected in the outcomes of optical conductivity $\sigma_1(\omega)$ (Fig. 2b), which have been derived from comprehensive spectroscopic ellipsometry measurements.

The $\sigma_1$ spectra of all NNO films exhibit common features including a peak $A$ at ~1.24 eV and a broad optical feature $C$ at ~3.05 eV. Based on the previous reports, peak $A$ is associated with the optical transition between occupied and unoccupied antibonding $e_g$* orbitals [43, 44], while feature $C$ arises from the transition between the O $2p$ orbital and the unoccupied $e_g$* band [43]. The $\sigma_1$ spectra of both NNO/LAO and NNO/LSAT are in line with previous reports [45, 46, 47, 48], while the spectra of NNO/STO exhibit notable deviations in optical characteristics that have yet to be explained. We observe the emergence of feature $B$ (~2.12 eV), and features $D$ (~3.75 eV), $E$ (~4.10 eV) and $F$ (~4.67 eV).

To uncover the origins of these unique optical features in the NNO/STO films, we conducted temperature-dependent ellipsometry measurements. Figs. 2c–e display the combination of temperature-dependent $\sigma_1$-spectra (from 0.6 to 5 eV) and transport measurements (0 eV) of NNO/STO samples with varying NNO film thickness. The optical conductivity varies with temperature, particularly in the 0-0.6 eV energy region. For all NNO/STO samples, there are two distinct peak features labeled $A$ (~0.78 eV) and $A'$ (~1.24 eV) in the $\sigma_1$-spectra measured at 77 K. These low-temperature optical features can be attributed to the onset of bond disproportionation in the low-temperature insulating phase [49, 50]. As the temperature increases and the insulator-to-metal transition occurs, these two-peak features merge into a single peak. The prominent feature $B$ (located at ~2.12 eV) observed in the 4.0 nm and 7.0 nm NNO/STO samples, absent for NNO films grown on LAO and LSAT (Fig. 2b), persists throughout the entire temperature range from 77 and 300 K (Fig. 2c-d). Specifically, as the film thickness continues to increase, feature $B$ becomes visible only at 77 K in the 12.0 nm thick NNO/STO sample (Fig. 2e) and this feature is suppressed at higher temperatures. The progressive suppression of feature $B$ with increasing NNO thickness implies that this unique optical feature, specific to NNO/STO, is likely attributed to interfacial effects occurring close to the NNO/STO interface.

In the context of higher photon energy (above 3.5 eV), the $\sigma_1$ spectra of NNO/STO samples with varying film thicknesses reveal two optical features, denoted as $D$ (~3.75 eV) and $F$ (~4.67 eV). Additionally, between features D and F, a shoulder labelled $E$ (~4.10 eV) is observed. Subsequent detailed discussions further attribute these optical features to the resonant excitonic effects originating from the STO substrate [51].

After analyzing the $\sigma_1$ spectrum, we proceed to discuss the variations in $N_{eff}$ caused by spectral weight. The metal-insulator transition can also be assessed by the comparison of the effective number of free carriers ($N_{eff}$) for different temperatures. To estimate $N_{eff}$, we supplemented the optical conductivity at 0-0.6 eV with transport measurements (inset of Figs. 2c-e). As it can be observed, for 4.0 nm and 7.0 nm NNO/STO, $N_{eff}$ slightly increases with decreasing temperature at high temperatures, continuing until below $T_{MI}$ (Fig. 2f). Subsequently, a significant decrease in $N_{eff}$ occurs due to the formation of the gap associated with the metal-insulator (MI) transition. For 12.0 nm NNO/STO, we can also determine the temperature range in which its metal insulation transition occurs. Therefore, the analysis of $N_{eff}$ is consistent with the transport measurements.

**Electronic states of NNO films as revealed by XAS**

To precisely determine the origin of optical features in NNO/STO and further analyze their temperature- and thickness-dependent behavior, O $K$-edge and Ni $L$-edge XAS are conducted for the respective samples. Fig. 3a compares the XAS spectra at the O $K$-edge (O $1s$—$2p$ transition) of the respective NNO films.

The prominent pre-edge feature labeled $a$, present in all NNO films at ~528.1 eV, is attributed to the Ni $3d$—O $2p$ orbital hybridization, as discussed in previous studies [29, 52, 53]. The broad and resonant feature labeled $b$ at a higher energy position of ~535.4 eV is ascribed to the orbital hybridization between Nd $5d/4f$ and O $2p$ orbitals [54, 55, 56]. Further in the photon energy region from ~537 to 547 eV, features $c$ and $d$ correspond to the hybridization between the Ni $4sp$/Nd $5sp$ and O $2p$ orbitals [54, 55, 56]. Despite the thorough consideration of the O $K$-edge absorption features, it is crucial to note a unique small absorption feature labeled as $a^*$ at ~530.5 eV, only observed in the NNO/STO sample, but absent in other NNO films. Feature $a^*$ is notably pronounced in the thinnest 4.0 nm-thick NNO/STO sample, suggesting the existence of a new unoccupied hybridized state (with O $1s$) at the NNO/STO interface.

Interestingly, the intensity of feature *a** (Figs. 3a and b) exhibits a thickness-dependent trend, mirroring that of optical feature *B* in the $\sigma_1$ spectra of NNO/STO samples with varying thicknesses, as detected using spectroscopic ellipsometry (Fig. 2e). The consistent thickness-dependent behavior observed in these two distinct experimental techniques leads to the deduction that features *a** and *B* likely share the same origin.

In previous studies involving heterointerfaces with STO as the substrate material, optical features similar to what we have observed as *a** were linked to the interfacial hybridization between the Ti 3*d* orbital of the STO substrate and the O 2*p* orbital of the oxide thin-film [7, 12, 57]. To further substantiate the evidence of interfacial hybridization at the NNO/STO interface, particularly in the case of the 4.0 nm-thick NNO/STO sample, we conducted polarization-dependent O *K*-edge spectra measurements. Fig. 3b displays the intensity variation of the O *K*-edge at 20° (out-of-plane E-field), 90° (in-plane E-field) polarizations, and at an intermediate angle of 40° (inset of Fig. 3b). Notably, the intensity of feature *a** peak at 20º (out-of-plane) polarization weakens with increasing polarization angle. The reduced absorption intensity for the in-plane polarization spectra suggests more out-of-plane empty states,  suggesting a stronger out-of-plane orbital hybridization between the Ti 3*d* orbital of the STO substrate and the O 2*p* orbital of the NNO film at the film-substrate interface.

Ni *L*-edge XAS measurements were also conducted for the NNO/STO samples with varying film thicknesses (Fig. S7). The shape and peak positions of both the $L_3$ and $L_2$ edges remain generally unchanged [58, 59, 60]. These features correspond to the Ni 2*p* $\rightarrow$ 3*d* transition [32, 54] present in Ni electronic states. To further confirm that the XAS measurements probe the interfacial effects at the NNO/STO interface, we also measured the Ni *L*-edge XAS of 12.0 nm and 14.0 nm-thick NNO/LAO samples (Fig. S8). In addition to the Ni $L_{3,2}$ edges attributed to the NNO layer, an additional feature is observed at ~850 eV for the 12.0 nm-thick NNO/LAO sample, attributed to the La $M_4$-edge feature belonging to the LAO substrate [37, 59]. However, the feature is no longer visible in the 14.0 nm-thick NNO/LAO sample, indicating that the probing depth of the soft XAS technique is ~12.0 nm. Taken together, these findings suggest that XAS for 4.0, 7.0 and 12.0 nm-thick NNO/STO samples include signals from their respective interfaces and the underlying hybridization between the STO substrate and NNO film. Thus, the observed feature $a^*$, notably prominent in the 4.0 nm-thick NNO/STO sample and diminishing with increasing the NNO film thickness, can be attributed to the existence of interfacial orbital hybridization at the NNO/STO interface. EDS maps (Fig. 1d) on the other hand confirmed this conclusion, the atomic mixing detected at the interface also enhanced the hybridization between atomic orbitals [61].

**Electronic band diagram and interfacial effects of STO on NNO film**

A combined analysis of both the XAS spectra and the $\sigma_1$ spectra enables us to construct an electronic band model for the 4.0 nm-thick NNO/STO sample in both metallic and insulating phases. The electronic band model schematic in Fig. 3c illustrates the unoccupied bands corresponding to the resonant absorption peaks observed in the O $K$-edge spectra (Figs. 3a and 3b). The energy differences for each band are derived from the peak positions in the optical conductivity spectra [62]. Since the XAS spectra are acquired at room temperature, they reflect the electronic structures in the high-temperature metallic state (top panel Fig. 3c).

The pre-peak of the O $K$-edge spectra consists of two distinct features: peak $a$ (~528.1 eV) and a relatively weaker shoulder feature at 529.2 eV. The broad peak feature $a$ in the XAS spectra can be attributed to the $e_g^*$ and O $2p$ unoccupied state. Above the Fermi surface, the unoccupied $e_g^*$ band close to the unoccupied segment of the $eg^*$ and O $2p$ band manifests itself as the shoulder above peak $a$ in the O $K$-edge spectra.

As previously mentioned, the energy position of feature $a^*$ at ~2.5 eV above peak $a$ (Fig. 3a) corresponds to a previously unidentified unoccupied hybridized state at the same energy level above the Fermi surface. The formation of this state is induced by the robust Ti $3d$—O $2p$ orbital hybridization at the interface between the O $2p$-orbitals of the NNO film and the Ti $3d$-orbitals of the STO substrate. Meanwhile, when considering feature B in the $\sigma_1$-spectrum (Fig. 2b) alongside the O $K$-edge spectra, it can be deduced to correspond to the optical transition from Ni $3d$ $t_{2g}^*$ bands to the newly-formed unoccupied hybridized state $a^*$ recorded by the XAS measurement. Note that the slight energy mismatch between the ellipsometry (~2.12 eV) and XAS data (~2.6 eV) can be attributed to differences in spectral resolution between these two techniques.

Having established the schematic band diagram for the metallic phase of the 4.0 nm-thick NNO/STO sample, we can extend our analysis to the system's low-temperature insulating state using the $\sigma_1$ spectra at 77 K, where features A and A' emerge (Fig. 2c-e). This emergence can be attributed to the onset of bond disproportionation in the low-temperature insulating phase [49, 50], resulting in the split of the Ni-O hybridized band. Based on the separation distance between features A and A', we can assign the unfilled segment of the Ni-O hybridized band to be located ~0.6 eV above the Fermi energy (bottom panel Fig. 3c).

Having established the band diagram of the 4.0 nm-NNO/STO based on the spectroscopic ellipsometry and XAS data and combining the results with detailed analysis, it can be seen more intuitively that the interfacial orbital hybridization between the Ti $3d$ belonging to the STO substrate and the O $2p$ orbitals of the NNO film results in the formation of the new unoccupied state above the Fermi surface which is manifested as peak $a*$ in the O $K$-edge XAS spectra.

**Excitonic effects from the STO substrate**

To further understand the excitonic effects originating from the STO substrate and verify the existence of Ti $3d$-O $2p$ hybridization at the NNO/STO interface, a temperature-dependent differential $\sigma_1$, denoted as $\Delta\sigma_1$ (where $\Delta\sigma_1(\omega,T)=\sigma_1(\omega,T)-\sigma_1(\omega,300K)$), analysis has been conducted for NNO/STO samples with varying NNO film thicknesses in the high photon energy regions above 3.5 eV (Figs. 3d–f). These regions have been previously reported for exhibiting significant effects on thin films due to resonant excitons from the STO substrate [51, 63]. In the $\Delta\sigma_1$ spectra, two visible temperature-dependent optical structures located at ~3.75 eV and ~4.67 eV have been observed across all NNO/STO samples. Intriguingly, these structures align with two exciton peaks [64] observed at the same photon energy positions on bulk STO (Fig. 3g), where they have been identified as Wannier and resonant excitons, respectively [51]. Furthermore, these features exhibit a similar temperature-dependent behavior to that observed in bulk STO, notably weakening significantly with increasing temperature. Previous studies have suggested that excitonic wave functions can propagate from the STO substrate to the thin-film via interfacial orbital hybridization [10, 64, 65]. Our experimental results indicate the extension of excitonic properties from the STO substrate into the NNO film.

While a similar temperature-dependent trend persists with increasing the NNO film thickness, the excitonic features observed in the $\Delta\sigma_1$ for both 7.0 nm-thick and 12.0 nm-thick NNO/STO samples gradually diminish, as seen in Figs. 3e–f. This consistency aligns with the experimental observation in the $\sigma_1$ spectra for the NNO/STO samples at respective thicknesses, where peaks $D$ and $F$ are located (Figs. 2c–e). Meanwhile, the $\Delta\sigma_1$ spectra for the 12.0 nm-thick NNO/LSAT sample display minimal temperature-dependent behavior. This discrepancy allows us to dismiss the notion that the formation of the high-energy optical feature is solely due to lattice strain effects. The presence of optical features within the photon energy range between ~3.6 and 4.7 eV in the NNO/STO samples underscores the significant impact of the STO substrate on the NNO film, attributed to the intrinsic electron-electron (e–e) and electron-hole (e-h) interactions within the system.

**DFT results of the NNO/STO interface**

To further investigate the interfacial interaction between the NNO film and the STO substrate and its substantial impact on the electronic and optical structures of the NNO/STO sample, we conducted first-principles density functional theory (DFT) calculation (Fig. 4a) [66, 67, 68, 69]. Figs. 4b–e display the partial density of states (PDOS) of the Ti, O and Ni components at the interface, while the hybridization between O $2p$ and the Ni/Ti $3d$ orbitals is schematically illustrated in Fig. 4f. When considering charge ordering in the NNO layer, two types of Ni ions emerge, characterized by the formal oxidation states of $Ni^{2+}$ and $Ni^{4+}$. These two Ni ions correspond to larger and smaller Ni-O octahedra size, labeled as $Ni_L$ and $Ni_S$, respectively. The $e_g$ orbitals of $Ni_L$ are occupied by two electrons with two empty states separated by a Mott-Hubbard gap. For the case of $Ni_S$ ions, the anti-bonding O $2p$ $e_g$ orbitals are unoccupied.

The difference in the energy gap between $Ni_L$ and $Ni_S$ ions indicates a site-selective Mott transition [70]. It is noteworthy that the spin polarization for $Ni_S$ shown in Fig. 4e vanishes in the $S$-type antiferromagnetic structure [71]. These $e_g^A$ (A means antibonding) state are at around 0~3 eV. The interfacial Ti $3d$ states shows significant DOS in the energy region between ~3 to 5 eV (Fig. 4b). Moreover, the O $2p$ states also has dominant contribution to the DOS between ~3 to 5 eV (Fig. 4c). While these states are absent from the Ni $3d$ PDOS (Figs. 4d and 4e). The dominant presence of both the Ti $3d$ and O $2p$ states are therefore clear indications that strong O $2p$-Ti $3d$ orbital hybridization is taking place. This leads to the formation of an interfacial state in this energy region. Thus, this phenomenon accounts for the appearance of feature B in the optical spectra, a characteristic unique to the NNO/STO system and consistently observed throughout the entire temperature range, as shown in Figs. 2c-e.

While a strong interaction is observed between the NNO film and the STO substrate, such a phenomenon is not replicated in the case of NNO on other substrates. For instance, in the case of NNO/LAO, where the Al $3p$ orbitals are situated at a relatively higher energy level, the interfacial hybridization between Al $3p$ and O $2p$ is significantly weaker, rendering any effects less detectable (Fig. S9). This is also the reason why the optical feature $B$ is absent for NNO films grown on other substrates, as illustrated in Fig. 2b.

**Conclusions**

In conclusion, our observation indicates that the NNO thin films grown on STO substrates exhibit a distinctive band structure compared to films grown on other substrates. This distinctiveness is characterized by the presence of an additional unoccupied band located above the Fermi level. Through a comprehensive investigation that comprises temperature-dependent spectroscopy ellipsometry, X-ray linear dichroism, and DFT calculations, we determine that the extra unoccupied band in the NNO/STO samples is a result of the interfacial orbital hybridization between Ti $3d$ orbitals of the STO substrate and O $2p$ orbitals of the NNO film. These findings highlight the significance of atomic orbital hybridization at oxide interfaces, providing valuable insights into the effects of these interface. The ability to design interface structures and manipulate atomic orbital interactions not only advances our fundamental understanding of complex oxide heterostructures, but also presents exciting opportunities for controlling the electronic structure of materials. Such control holds promising implications for various applications, ranging from electronic devices to energy storage systems.

**METHODS**

**Sample preparation and characterization.** NNO films with thickness varying from 4 to 12 nm were deposited on diverse (001)-oriented single crystal substrates using PLD. Throughout the deposition process, the substrate was hold at 675 °C, with an oxygen partial pressure of 300 mTorr. The 248 nm laser fluence was ~ 2 $J/cm^2$, operating at a repetition rate of 5 Hz. After deposition, the oxygen partial pressure was increased to 10 kPa, and the samples were cooled to room temperature at a ramping rate of 25°C/min. High resolution XRD measurements were conducted with a Rigaku Smart Lab XE instrument equipped with a rotating anode Cu Kα source, Ge (220) 2-bounce incident monochromator, and HyPix-3000 2-dimensional detector. The analysis of structural and chemical composition involved HAADF imaging and EDS within a STEM. To prepare the TEM sample, a dual beam Helios instrument was employed which combines both focused ion beam (FIB) and scanning electron microscopy (SEM). Initially, a cross-sectional lamella was extracted from thin film sample through FIB milling, gradually thinning it to approximately 200 nm at 30 kV. The sample was further reduced to about 50 nm at 5 kV before its final polishing at 2 kV. For HAADF imaging, a Thermo Fisher Scientific Spectra Ultra S/TEM equipped with an aberration corrector and a monochromator was used, with a convergence angle of 30 mrad and a collection angle ranging from 60 to 180 mrad. Atomic-scale EDS acquisition and analysis was performed using an advanced EDS detector (Ultra-X) within the STEM with a solid angle of 4.45 srad and operating at a probe current of 42 pA in Velox software. In-plane transport measurements were performed using a 9 T PPMS (Physical Properties Measurement System, Quantum Design) at temperatures ranging from 10 to 300 K at a cooling/warming rate of 3 K/min.

**Spectroscopic ellipsometry measurements.** Spectroscopic ellipsometry measurements are conducted using a custom-made Variable Angle Spectroscopic Ellipsometer (VASE) of J. A. Woollam Co., Inc with the photon energy range of 0.60 – 4.80 eV at incident angles of 70° with respect to the plane normal. Ellipsometry parameters Ψ (the ratio between the amplitude of $p$ and $s$-polarized reflected light) and Δ (the phase difference between $p$ and $s$-polarized reflected light) in a high vacuum chamber with a base pressure of $1\times10^{-9}$ mbar at the respective temperature. The substrate layers (bulk LAO, LSAT or STO) have also been measured under the same conditions. The optical conductivity, $\sigma_1$, of the respective NNO thin-film layers have been extracted from parameters Ψ and Δ utilizing an air/NNO/Substrate multilayer model, where NNO film comprises a homogeneously uniform medium and a composite heterointerface component.

**XAS.** The O $K$-edge absorption spectra in the energy range 520–580 eV and Ni $L$-edge absorption spectra in the energy range 845–880 eV were obtained using linearly polarized X-ray absorption spectroscopy at the Surface, Interface and Nanostructure Science (SINS) beamline at Singapore Synchrotron Light Source (SSLS), using the Total Electron Yield detection mode. The experimental was conducted in an ultrahigh vacuum chamber with a base pressure of $1\times10^{-10}$ mbar. The spectra were normalized to the integrated intensity between 565 and 580 eV for the O $K$-edge spectra and between 870 and 880 eV for the Ni $L$-edge spectra where the tail end of the spectra is located.

**DFT.** The first-principles calculations were performed with the DFT as implemented in the SIESTA code[72, 73]. We use the PBEsol functional which is optimized for predicting the structures of solid structures. The pseudopotentials are from the Pseudo-dojo dataset[74]. A Hubbard U correction[75] is used to get better description of the Ni 3d electronic structure, with an U(Ni)=2 eV. We use a double-zeta-polarized basis set for the electron wavefunction. The k-point mesh is set to 6×6×1 for the heterostructure. A structure with 4 u.c. of NNO and 4 u.c. of STO, with NdO-TiO$_2$ layers at the interface, and a vacuum layer of 20 Angstrom at both end. In each layer, a √2×√2-unit cell structure is taken so that the rotations and the breathing distortions of the oxygen octahedra are allowed. A ferromagnetic spin configuration is assumed in the NNO structure instead of the ground state of the S-type antiferromagnetic structure to reduce the computational complexity. The in-plane lattice parameters are constrained to those of the substrate (3.905 Å for STO). The structures are relaxed until the forces are less than $3 \times 10^{-3}$ eV/Å, and the stresses less than 1 GPa.

## DATA AVAILABILITY

The data that support the findings of this study are available from the corresponding author upon reasonable request.

**ACKNOWLEDGMENTS**

The authors would like to acknowledge the Singapore Synchrotron Light Source for providing the facility necessary for conducting the research. The Laboratory is a National Research Infrastructure under the National Research Foundation, Singapore. Any opinions, findings, and conclusions or recommendations expressed in this material are those of the author(s) and do not reflect the views of National Research Foundation, Singapore. This work was supported by National Natural Science Foundation of China (Grant Nos. 52172271, 12374378, 52307026); the National Key R&D Program of China (Grant No. 2022YFE03150200); Shanghai Science and Technology Innovation Program (Grant No. 22511100200); the Strategic Priority Research Program of the Chinese Academy of Sciences (Grant No. XDB25000000). XRD and STEM measurements along with the corresponding analysis and manuscript writing were supported by the U.S. Department of Energy (DOE), Office of Science, Basic Energy Sciences, Division of Materials Sciences and Engineering, Synthesis and Processing Science Program, under Award #10122. C.S.T acknowledges the support from the NUS Emerging Scientist Fellowship. X.H. acknowledge financial support from F.R.S.- FNRS Belgium through the PDR project PROMOSPAN (Grant No. T.0107.20).

**CONTRIBUTIONS**

X.Y. conceived the project. L.W. synthesized the thin films. Y.L. conducted the in-plane transport experiments and L.W. performed the data analysis. M.B. carried out the XRD measurements. K.R.K. conducted the STEM measurements and data analysis under the supervision of Y.D. C.S.T. performed the XAS experiments and M.C. analyzed the data. M.S., C.S.T., X.L., H.L. performed the SE experiments and analyzed the data. X.H. did the DFT simulations. M.C., C.S.T., X.H., L.W., and X.Y. wrote the manuscript, with input from all the authors.

**DECLARATION OF INTERESTS**

The authors declare no competing interests.

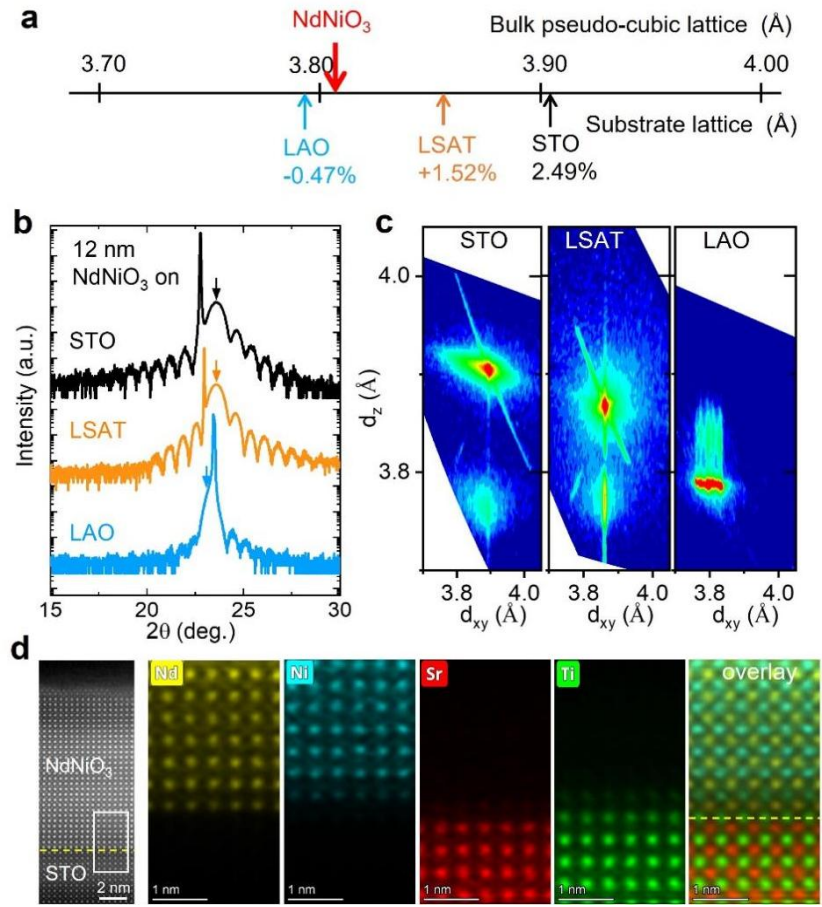

**Fig. 1. Synthesis of NNO films and structural characterization.** (a) Lattice mismatch between bulk NNO and the substrates used in this study. (b) XRD $\theta$–$2\theta$ scans near the (001) peak of NNO films on STO, LSAT and LAO substrates. NNO films peaks are marked by arrows. (c) Reciprocal space mapping (RSM) for 12.0 nm thick NNO on STO, LSAT and LAO substrates. (d) Cross-sectional STEM image (left) of the NNO/STO interface viewed along the [100]-direction of STO. Atomic-resolution EDX mapping (right) of the NNO/STO interface. Yellow, blue, red, and green respectively represent the Nd, Ni, Sr, and Ti atoms.

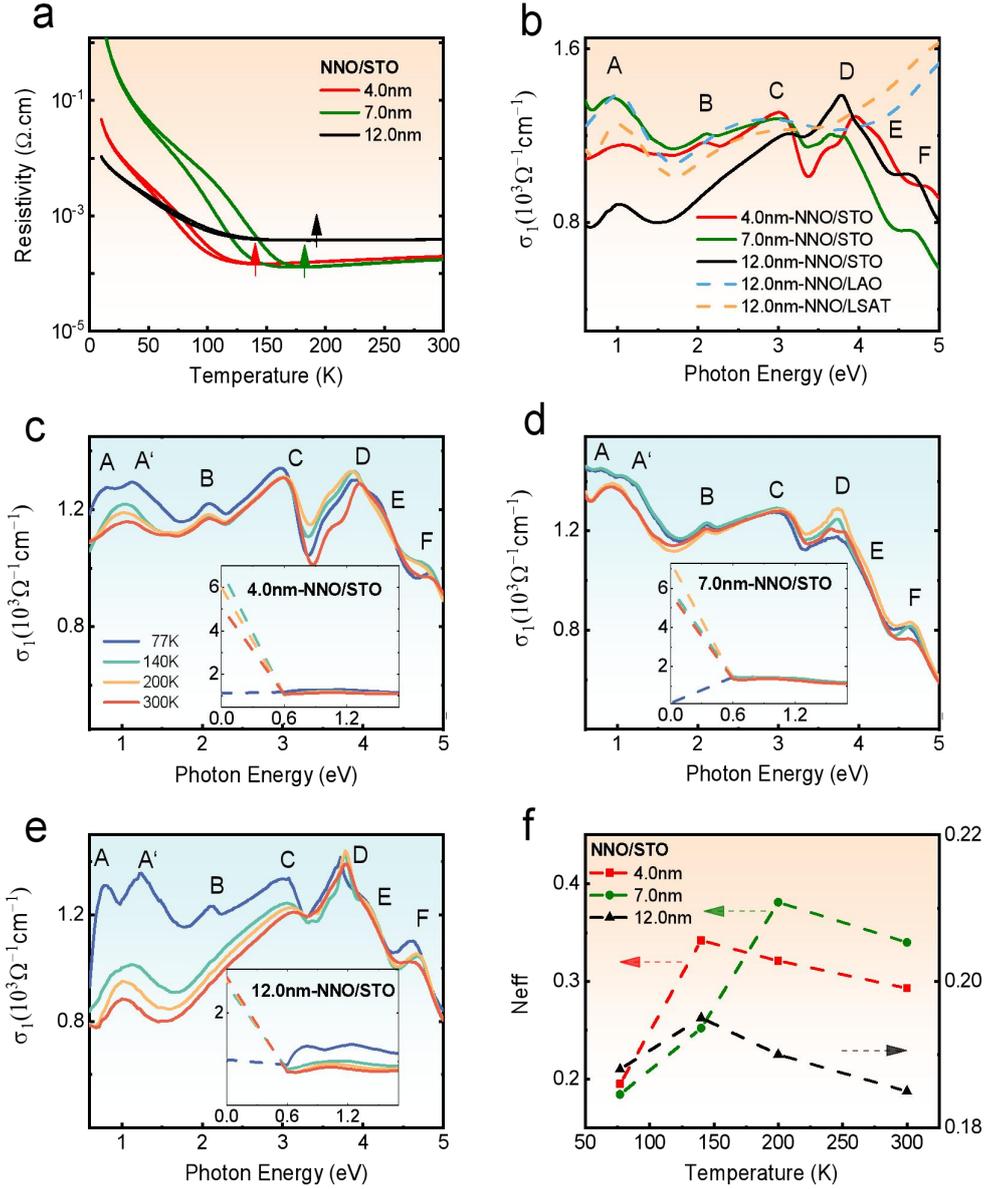

**Fig. 2. Temperature-dependent electrical and optical properties of NNO films.** (a) Temperature-dependent resistivity of the NNO/STO samples with different film thicknesses. The inset shows the 12.0 nm thick NNO films on various substrates. Arrows indicate the temperature at which metal-insulator transition takes place in each thin-film sample. (b) Room temperature optical conductivity, $\sigma_1$, of the respective NNO films. Temperature-dependent $\sigma_1$ spectra of (c) 4.0 nm, (d) 7.0 nm, and (e) 12.0 nm thick NNO films grown on STO at temperatures ranging from 77 to 300 K. The insets in the (c-e) are the temperature-dependent $\sigma_1$ spectra for 0-1.7 eV of respective NNO/STO thin films. The $\sigma_1$ at 0 eV is estimated from dc conductivity in Fig. 2a. (f) The number of effective charge ( $N_{eff}$ ) as a function of temperature for 0 – 1.7 eV.

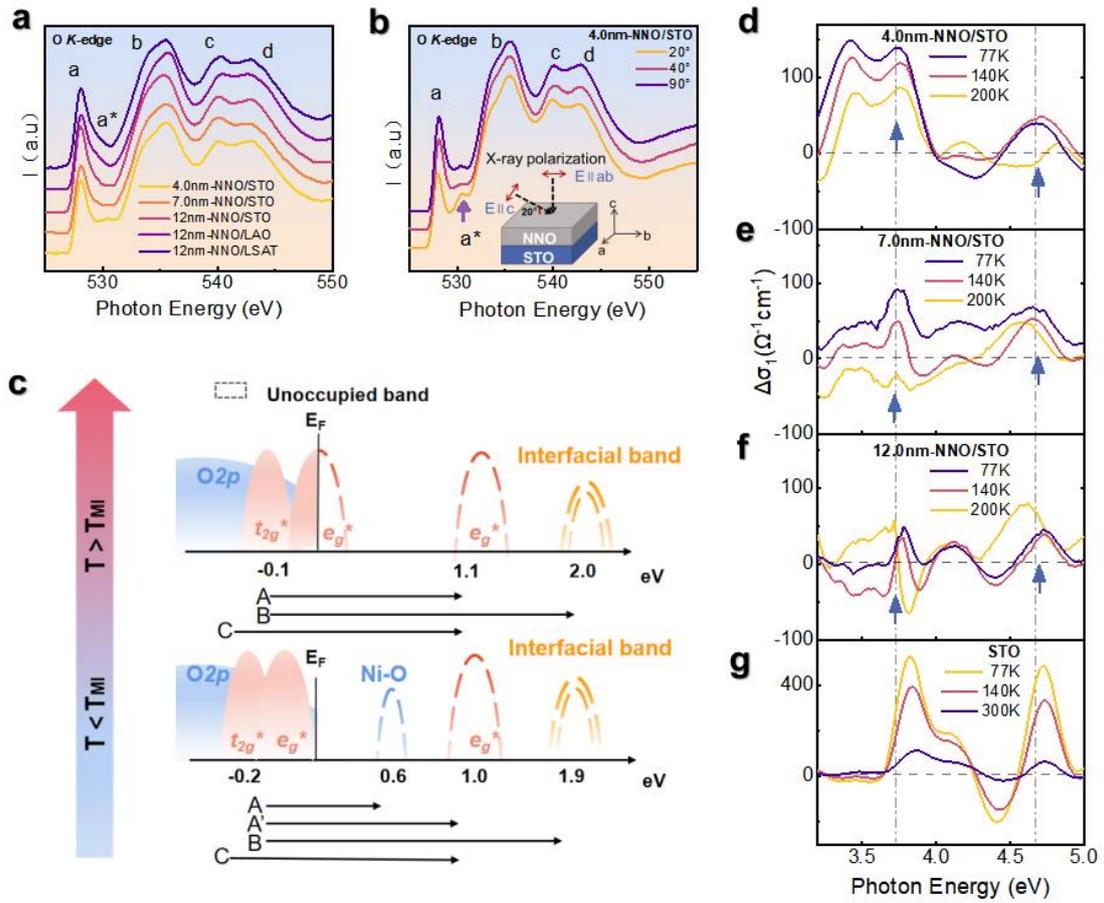

**Fig. 3. O *K*-edge XAS of NNO film samples and energy band schematics**. (a) O *K*-edge XAS of the NNO film samples. (b) Polarization-dependent O *K*-edge XAS of the 4.0 nm-thick NNO/STO sample. (c) Energy band schematic of the 4.0 nm-thick NNO/STO sample in the high-temperature metallic state (top) and low-temperature insulating state (bottom). Differential optical conductivity, $\Delta\sigma_1$ (where $\Delta\sigma_1(\omega,T)=\sigma_1(\omega,T)-\sigma_1(\omega,300\ K)$), of (d) 4.0 nm-NNO/STO, (e) 7.0 nm-NNO/STO, (f) 12.0 nm-NNO/STO, and in comparison with (g) bare STO substrate.

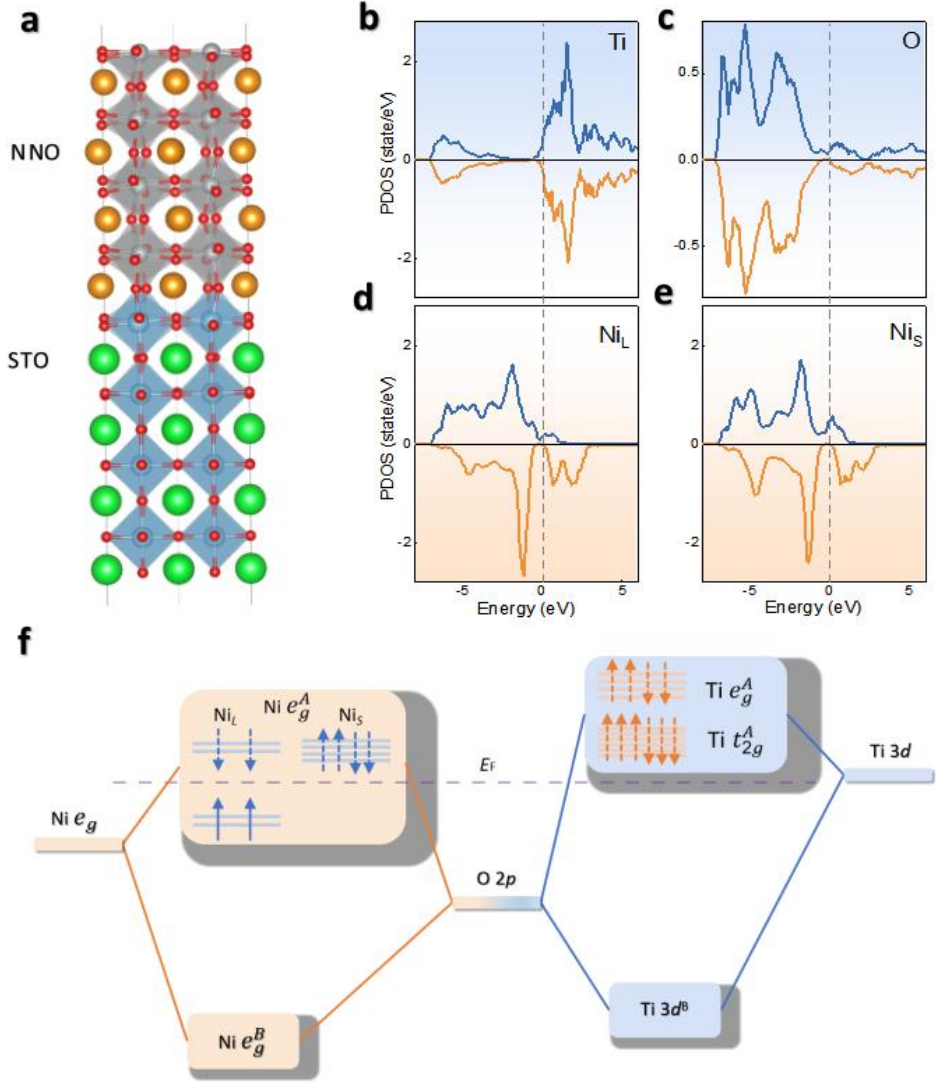

**Fig. 4. DFT results.** (a) NNO/STO superlattice structure modelled for the density of states projected to the (b) Ti, (c) O, and (d) $Ni_L$, and (e) $Ni_S$ atoms at the interface layers in the phase with Ni-O octahedral breathing distortion. (f) The schematic figure of the hybridization between the Ni, O, and Ti orbitals. The Ni $t_{2g}$ orbitals have not been shown for simplicity. The detailed energy splitting of the bonding state of the Ni $e_g$ and O $2p$ orbitals, and the bonding state between Ti $3d$ and O $2p$ orbitals have been displayed. Meanwhile, the other states are shown only in non-detailed way. Superscripts $A$ and $B$ labels indicate the anti-bonding and bonding states, respectively.